\pdfoutput=1

\documentclass[%
 reprint,
superscriptaddress,
 amsmath,amssymb,
 aps,
prb,
nofootinbib,
]{revtex4-2}

\usepackage{tabularx}
\usepackage{hyperref} 
\hypersetup{colorlinks=true, citecolor=blue, urlcolor=blue, linkcolor=blue} \usepackage{siunitx}
\usepackage{graphicx}
\usepackage{bm}

\begin{document}


\title{Thermodynamic stability of Li-B-C compounds from first principles}

\author{Saba Kharabadze} 
\affiliation{
  Department of Physics, Applied Physics and Astronomy,\\ Binghamton University,         State University of New York, \\PO Box 6000, Binghamton, New York 13902-6000, USA.
}

\author{Maxwell Meyers} 
\affiliation{
  Department of Physics, Applied Physics and Astronomy,\\ Binghamton University,         State University of New York, \\PO Box 6000, Binghamton, New York 13902-6000, USA.
}

\author{Charlsey R. Tomassetti}
\affiliation{
  Department of Physics, Applied Physics and Astronomy,\\ Binghamton University,         State University of New York, \\PO Box 6000, Binghamton, New York 13902-6000, USA.
}

\author{Elena R. Margine} 
\affiliation{
  Department of Physics, Applied Physics and Astronomy,\\ Binghamton University,         State University of New York, \\PO Box 6000, Binghamton, New York 13902-6000, USA.
}
\author{Igor I. Mazin}
\affiliation{
Department of Physics and Astronomy, George Mason University, Fairfax, Virginia 22030, USA\\
and Quantum Science and Engineering Center, George Mason University, Fairfax, Virginia 22030, USA
}

\author{Aleksey N. Kolmogorov\textit{$^{\ast}$}}
\affiliation{
  Department of Physics, Applied Physics and Astronomy,\\ Binghamton University,         State University of New York, \\PO Box 6000, Binghamton, New York 13902-6000, USA.
}
%

\begin{abstract} Prediction of high-$T_{\rm{c}}$ superconductivity in hole-doped Li$_x$BC two decades ago has brought about an extensive effort to synthesize new materials with honeycomb B-C layers, but the thermodynamic stability of Li-B-C compounds remains largely unexplored. In this study, we use density functional theory to characterize well-established and recently reported Li-B-C phases. Our calculation of the Li chemical potential in Li$_x$BC helps estimate the ($T$,$P$) conditions required for delithiation of the LiBC parent material, while examination of B-C phases helps rationalize the observation of metastable BC$_3$ polymorphs with honeycomb and diamond-like morphologies. At the same time, we demonstrate that recently reported BC$_3$, LiBC$_3$, and Li$_2$B$_2$C phases with new crystal structures are both dynamically and thermodynamically unstable. With a combination of evolutionary optimization and rational design, we identify considerably more natural and favorable Li$_2$B$_2$C configurations that, nevertheless, remain above the thermodynamic stability threshold.

\end{abstract}

\maketitle

\footnotetext{$^{\ast}$~Corresponding author: kolmogorov@binghamton.edu}




%



\section{Introduction} 

The combination of low atomic mass and strong covalent bonding makes binary and ternary Li-B-C compounds suitable for a variety of practical applications. The B$_{4-6.5}$C ceramic with outstanding Vickers hardness and high thermal stability has found uses as a protective material~\cite{thevenot1990}. Li-intercalated graphite compounds with the 372 mAh/g theoretical specific capacity have served as anodes in commercial Li-ion batteries~\cite{asenbauer2020}. The presence of high-frequency phonon modes and hole-doped electronic states makes Li-B-C materials with honeycomb layers particularly promising conventional superconductors with high critical temperatures ($T_{\rm{c}}$). For example, our ab initio re-examination of a recently synthesized LiB has indicated that the material’s $T_{\rm{c}}$ could exceed 32 K~\cite{ak48}. One of the most tantalizing predictions by Rosner et al. in 2002~\cite{Rosner2002} was the possibility of obtaining Li$_{x\approx{0.5}}$BC superconductors that could operate at temperatures above liquid nitrogen. However, the following synthesis and detailed characterization of the targeted delithiated Li$_x$BC material revealed no signs of superconductivity~\cite{Zhao2003, fogg2003, Fogg2006, Kalkan2019}. The search for new Li-B-C materials has continued in recent years and resulted in reports of new Li$_2$B$_2$C~\cite{pavlyuk2015}, BC$_3$~\cite{milashius2018}, and LiBC$_3$~\cite{milashius2017} phases with unique crystal structures.

\begin{figure}[t!] 
  \includegraphics[width=0.45\textwidth]{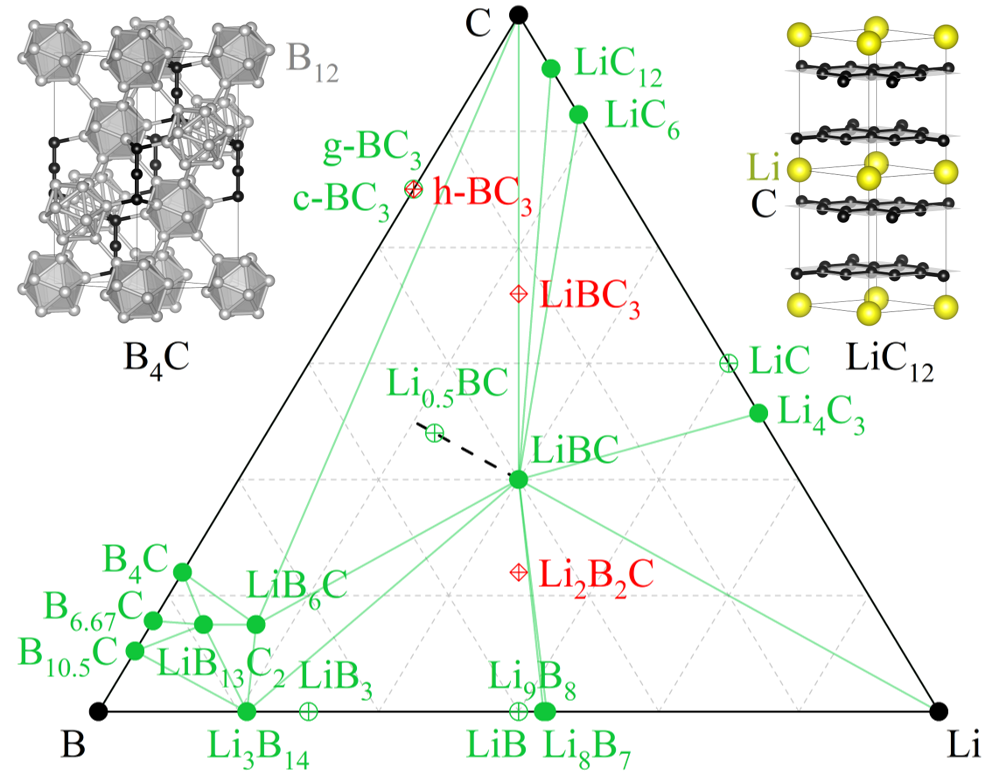} \caption{ 
  Convex hull of Li-B-C compounds determined with DFT (optB86b-vdW) calculations at $T=0$ K. Observed compounds found to be stable and metastable in this approximation are shown with solid and crossed green circles, respectively. Previously reported phases determined in this study to be unstable are displayed with crossed red circles. The dashed black line shows the observed stability range of Li$_x$BC phases. The morphology of the B$_4$C compound is shown with a simplified ordered metastable structure.}
\label{fig:all} \end{figure}

Key Li-B-C materials observed under ambient conditions are summarized in Fig.~\ref{fig:all}. The Li-B binary features Li$_3$B$_{14}$ and LiB$_3$ compounds with intercalated B frameworks~\cite{ak28}, an unusual LiB$_{x\approx 0.9}$ compound with linear B chains~\cite{worle2000,ak09}, and a predicted LiB with B layers synthesized  via cold compression  and annealed to 1 bar~\cite{ak08,ak30}. The Li-C system includes LiC$_{6n}$ ($n=1,2$) phases of Li-intercalated graphite, LiC with C$_2$ dimers, and Li$_4$C$_3$ with C$_3$ trimers~\cite{dresselhaus2002,sangster2007,lin2015}. The B-C binary contains B$_{4-6.5}$C related to pure B phases and different BC$_3$ polymorphs with 2D and 3D B-C frameworks~\cite{rogl2014,kouvetakis1986,zinin2007}. The well-established ternary compounds are LiB$_{13}$C$_2$ and LiB$_6$C with linked B$_{12}$ icosahedra~\cite{vojteer2006} and the layered LiBC~\cite{worle1995libc} with its delithiated Li$_x$BC derivatives ($x>0.38$)~\cite{fogg2003,Zhao2003,Fogg2006}. While the three binary systems have been the subject of several comprehensive ab initio modeling studies~\cite{ak09,lin2015,jay2019}, the thermodynamic stability of the Li-B-C ternary compounds has not yet been investigated systematically.

In this density functional theory (DFT) study, we focus on analyzing  the stability of published structural models for Li-B-C compounds. Firstly, we evaluate temperatures and Li$_2^{\textrm{gas}}$ vapor pressures needed to trigger the delithiation of LiBC. The constructed phase diagram for this process is shown to be consistent with the typical synthesis conditions used in experiments~\cite{Zhao2003,Fogg2006}. Secondly, we calculate the formation energies of B-C phases across a wide composition range to rationalize the observation of B-rich and C-rich materials. The observed BC$_3$ phases with layered~\cite{kouvetakis1986} or diamond-like~\cite{zinin2007} morphologies are confirmed to be only metastable~\cite{rogl2014}, which explains the need for the employed precursor-based synthesis routes. On the other hand, we show that the newest BC$_3$ polymorph, h-BC$_3$, comprised of alternating BC and C layers~\cite{milashius2018}, is highly unstable thermodynamically and appears to be an unlikely product of reaction between elemental B and C. Thirdly, we determine that the reported LiBC$_3$ phase in the form of intercalated h-BC$_3$~\cite{milashius2017} is a similarly unstable configuration. Finally, we show a number of unnatural features in the proposed structural model for the synthesized Li$_2$B$_2$C~\cite{pavlyuk2015}. We identify considerably more favorable configurations that remain only metastable. Our findings reveal an incomplete knowledge of the Li-B-C ternary and the need for further experimental exploration of this intriguing materials system.

\section{Methodology}

All DFT calculations were performed with the Vienna ab initio simulation package ({\small VASP})~\cite{VASP,VASP2,VASP3,VASP4}. The energy cutoff of 500 eV and dense ($\Delta k \sim 0.02$ $\si{\angstrom}^{-1}$) Monkhorst-Pack $k$-point meshes \cite{Monkhorst1976} ensured good numerical convergence of relative energies. All structures were fully relaxed with the EDIFFG=-0.005 tolerance, which led to the convergence of energies, forces, and stresses to typically below $\sim 1$ meV/atom, 0.005 eV/$\si{\angstrom}$, and 1 kB, respectively. Unless specified otherwise, we used the nonlocal van der Waals (vdW) functional optB86b-vdW~\cite{optB86b}. Select phases were examined with the generalized gradient approximation-based Perdue-Burke-Ernzerhof (PBE) exchange-correlation functional~\cite{PBE, Langreth1983} or within the local density approximation (LDA)~\cite{LDA1,LDA2}. The vdW-corrected functional was employed to account for dispersive interactions that play an important role for structural and bonding properties in layered Li-B-C materials. For instance, it allowed us to show that the disagreements between previously observed and our calculated interlayer distances in recently reported h-BC$_3$ and h-LiBC$_3$ polymorphs exceed the expected DFT errors. Formation energies of Li-C binary compounds have also been previously shown to be particularly sensitive to the systematic DFT errors~\cite{lin2015}. In our optB86b-vdW calculations, LiC is metastable at low temperatures but becomes thermodynamically stable around 300 K upon inclusion of the vibrational entropy. LiB$_3$ is also expected to stabilize at high temperatures~\cite{ak48}.

Global structure searches at specific Li-B-C compositions relied on an evolutionary algorithm implemented in the module for ab initio structure evolution {\small MAISE} \cite{ak41}. Populations of 16-20 structures with up to 20 atoms were generated randomly and evolved with our standard mutation and crossover operations \cite{ak41} for 20-50 generations. The thermodynamic corrections due to vibrational entropy were evaluated within the finite displacement method implemented in PHONOPY~\cite{Togo2015}. We used supercells with at least 80 atoms and applied 0.1 \AA\ displacements within the harmonic approximation. While the zero point energy (ZPE) is known to have a sizable magnitude in light materials, ranging from 40 meV/atom in bcc-Li to 171 meV/atom in graphite in our calculations, Table S1 shows that it canceled out effectively in the evaluation of formation energies and did not change the stability ordering for any of the considered phases. Since some of the Li-B-C structures were too large for phonon calculations or found to be dynamically unstable, we used energies in all the relative stability plots but showed values corrected with the ZPE and the vibrational entropy term at $T=600$~K for key materials in Table S1. We also examined the importance of anharmonic effects at high temperatures using the quasi-harmonic approximation (QHA). Detailed structural information for relevant phases considered in this study is given in the Supporting Information.

\section{Results and Discussion}

\begin{figure}[t!] 
\includegraphics[width=0.45\textwidth]{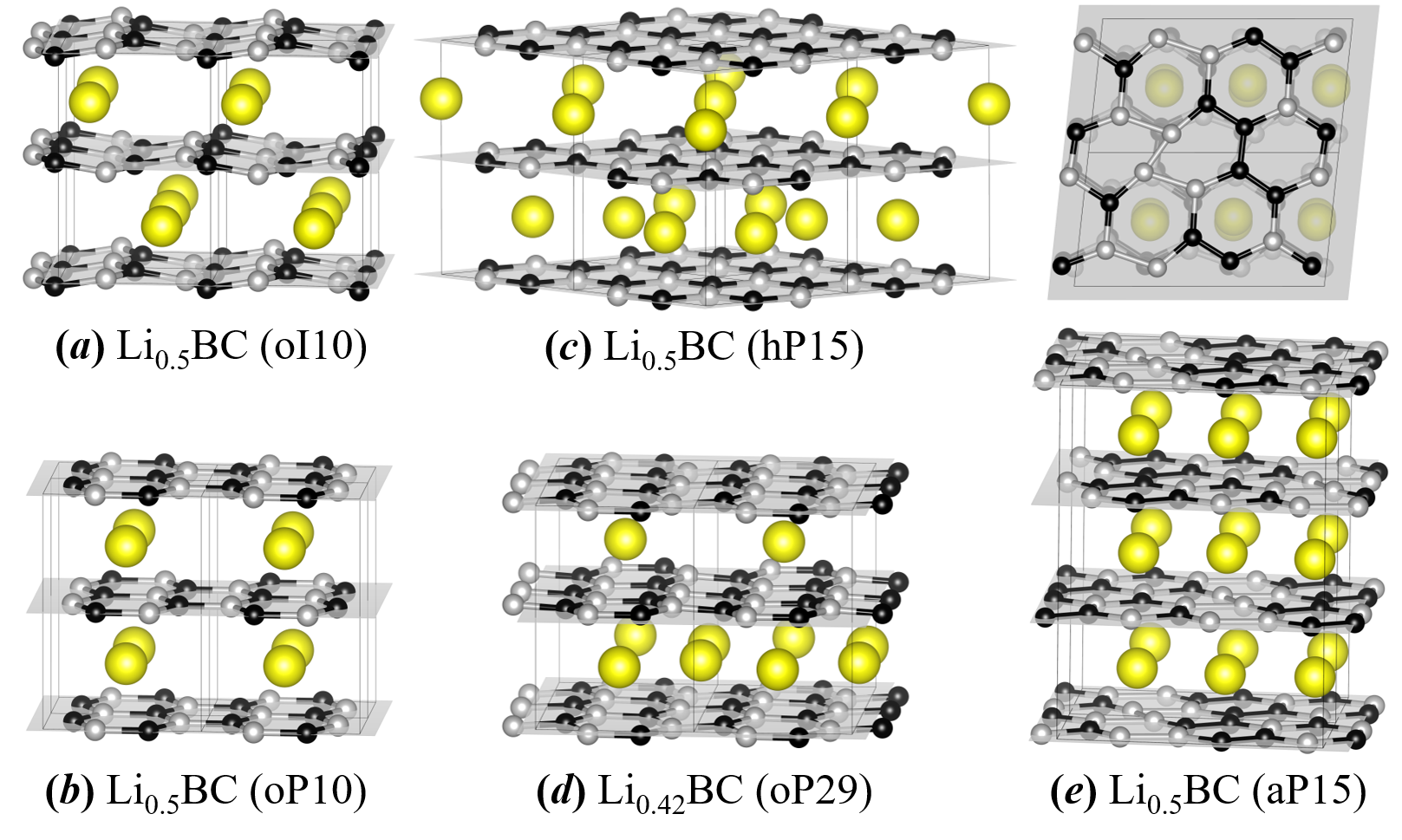} 
\caption{Simulated structural Li$_x$BC models at or near $x=0.5$. Phases (a-d) have uniformly ordered BC honeycomb layers and different populations of Li sites. The most stable aP15-Li$_{0.5}$BC polymorph (e) found in our evolutionary searches has a distorted BC network with C-C and B-B bonds.}
\label{fig:LixBC-pic} \end{figure}

\subsection{Li$_x$BC}
Wörle et al. first observed the formation of semiconducting LiBC in 1995~\cite{worle1995libc}. Particular interest in the layered compound was spurred by early 2000’s computational studies suggesting the material’s potential for high-$T_{\rm{c}}$ superconductivity in a hole-doped form~\cite{Ravindran2001, Rosner2002}. In the following few years, a number of different methods were used to obtain phases with reduced Li content, from delithiation of LiBC via vacuum annealing~\cite{Bharathi2002, Zhao2003} or oxidation in organic solvents~\cite{fogg2003} to direct synthesis of Li$_x$BC by varying the starting fluxes~\cite{Souptel2003,Nakamori2003}. The resulting compounds were obtained in a wide range of $x$, but those in the low Li regime ($x$ $<$ 0.5) were difficult to characterize, being highly disordered, multiphasic, amorphous, or with significant impurities~\cite{Bharathi2002, fogg2003, Souptel2003}. The low crystallinity of these Li-poor samples pointed towards a thermodynamic limit to the reduction of Li in LiBC. A later experimental study by Fogg et al.~\cite{Fogg2006} reported several critical concentrations where severe alterations in the LiBC structure were observed, including the swapping of B/C sites with the consequent formation of C-C and B-B bonds below $x \approx 0.55$ and the expulsion of B below $x \approx 0.45$. Detailed x-ray and neutron powder diffraction characterization of the obtained samples with $x=0.16-0.36$ nominal compositions indicated the formation of C-rich ternary and B-rich binary phases, e.g., Li$_{0.21(1)}$B$_{0.73(1)}$C$_{1.27(1)}$ and B$_{13}$C$_2$ phases at $x=0.16$. The idea of an upper limit to LiBC delithiation is further supported with DFT calculations~\cite{Fogg2006} and by the most recent findings by Kalkan et al.~\cite{Kalkan2019} who synthesized samples with Daumas-Hérold-type domains of Li$_x$BC in the 0.43 $\leq$ $x$ $\leq$ 0.85 range. 

The successful Li$_x$BC synthesis work has been followed by detailed measurements of materials' various properties~\cite{Zhao2003, fogg2003, Fogg2006, Kalkan2019, Xu2011}. Unfortunately, no superconductivity in Li$_x$BC has been detected in any of these studies, although Li reduction in LiBC has been shown to increase conductivity~\cite{Zhao2003, Fogg2006, Kalkan2019}. While ab initio studies of LiBC and its derivatives have examined the materials' bonding, superconducting, electrochemical, and other properties~\cite{An2003,Lee2003,Lebegue2004, Fogg2006,Liu2006,Xu2011,li2018}, the thermodynamics of the Li$_x$BC formation appears to be not fully explored.

\begin{figure}[t!] 
\includegraphics[width=0.45\textwidth]{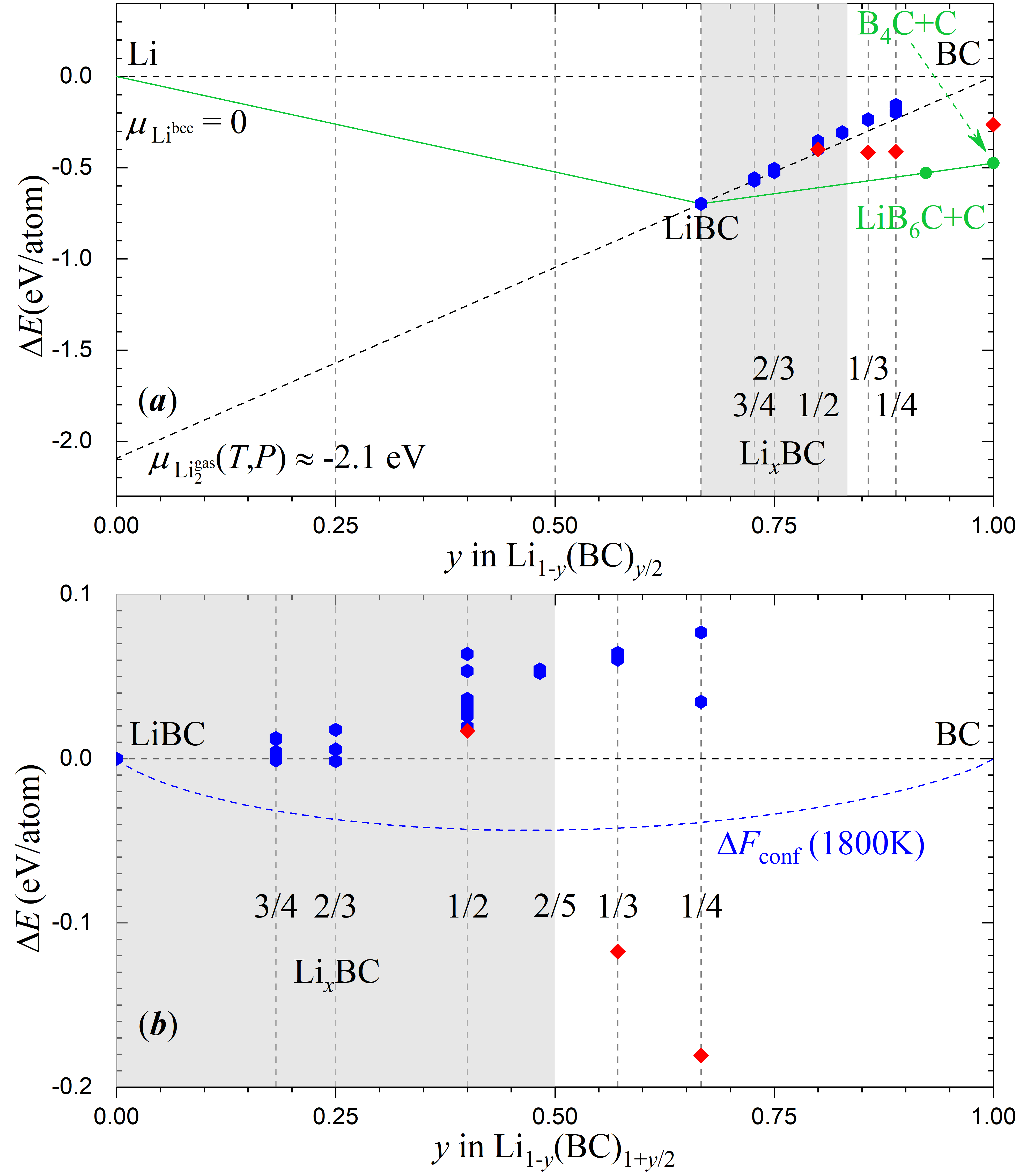}
\caption{Relative energies of Li$_x$BC pseudobinary phases referenced to (a) bcc-Li and layered BC or (b) LiBC and layered BC. The blue hexagons and red diamonds denote phases with honeycomb BC layers and general morphologies found with evolutionary searches, respectively. The green lines define the convex hull. The blue dashed line in (b) shows the configurational entropy contribution to the free energy for Li$_x$BC at 1800 K. The shaded region marks experimentally observed compositions.}   
\label{fig:LixBC-dat} \end{figure}

We began our investigation by screening the $0.25\leq x \leq 0.75$ pseudobinary range for favorable configurations using two complementing approaches. First, we systematically scanned all possible decorations of Li sites in LiBC supercells with up to 18 atoms and optimized the candidate structures Fig.~\ref{fig:LixBC-pic}. The formation energies of the best layered phases ended up well above the convex hull at $T=0$ K defined by combination of LiBC and the mixture of the pure C and LiB$_6$C materials (Fig.~\ref{fig:LixBC-dat}(a)) but fairly close to the line defined by the observed fully occupied and hypothetical fully delithiated compounds with the B-C frameworks (Fig.~\ref{fig:LixBC-dat}(b)). Different population of Li sites was found to disperse the energy by a sizable 45 meV/atom. At $x=0.5$, for example, structures with uniform distributions of Li atoms that broke the hexagonal symmetry and led to a noticeable buckling of the BC layers (Fig.~\ref{fig:LixBC-pic}(a,b)) were more favorable by at least 4.5 meV/atom than the hexagonal configuration with an uneven 2:1 population of sites in adjacent layers (Fig.~\ref{fig:LixBC-pic}(c)). Nevertheless, the kinetics of the deintercalation process may favor unequal populations as shown with the off-stoichiometry Li$_{0.42}$BC prototype (Fig.~\ref{fig:LixBC-pic}(d)) that approximates the experimentally observed decoration~\cite{Kalkan2019}.

Second, we performed evolutionary searches starting from random structures to explore configurations beyond the familiar layered morphologies. While the global optimization runs found the honeycomb frameworks to be lowest-energy minima for $x>0.5$, different motifs started to emerge at lower Li concentrations. At $x=0.5$, the low-symmetry structure with swapped B and C atoms causing a severe hexagon distortion at B sites (Fig.~\ref{fig:LixBC-pic}(d)), a pattern similar to the one observed in previous DFT simulations~\cite{Fogg2006}, turned out to be slightly favored, by 2 meV/atom, over all considered LiBC derivatives. At smaller $x$ = 1/3, 1/4, and 0 values, 3D frameworks comprised of sp$^2$ and/or sp$^3$ sites became dominant and lowered the energy by over 100 meV/atom. These findings are in line with the experimental observations regarding the delithiation limit, the defect types, and the lack of Li$_x$BC stability at $T=0$ K.

In order to account for the temperature-dependent factors determining the thermodynamics of the delithiation process, we evaluated the entropic terms in the Gibbs free energy for relevant Li-containing solid and gas phases. The configurational entropy of Li in disordered Li$_x$BC can be approximated assuming equiprobable occurrence of different decorations. The $\Delta F_{\textrm{conf}}=kT[x\ln(x)+(1-x)\ln(1-x)]/(2+x)$ contribution factoring in a variable number of Li atoms per formula unit is shown as a function of $y=(1-x)/(1+x/2)$ in Fig.~\ref{fig:LixBC-dat}(b). The correction brings the free energy of intermediate phases below the LiBC$\leftrightarrow$BC tie-line but is insufficient to stabilize them with respect to the convex hull even at 1800 K. The vibrational contribution evaluated within the harmonic approximation has a similar effect on Li$_x$BC stability. For example, the free energy distances to the LiBC$\leftrightarrow$(C+B$_4$C) tie-line at 1800 K for the ordered hP8-Li$_{2/3}$BC and oI10-Li$_{0.5}$BC changed from +118 to +99 meV/atom and from +210 to +175 meV/atom, respectively.

An unsurprising conclusion is that the deintercalation must be governed by the substantial entropy Li gains in its diatomic gas state. This contribution can be evaluated using calculated quantities for the vibrational, rotational, and translational degrees of freedom within the ideal diatomic gas model~\cite{chem-potential-textbook} (see Supporting Information). Fig.~\ref{fig:LixBC-dat}(a) illustrates that for LiBC to become thermodynamically unstable, the chemical potential of Li$_2^{\textrm{gas}}$ has to drop below a linearly extrapolated free energy value between the free energies of Li$_x$BC and LiBC for at least one $x$. Since ab initio computation of the configurational and vibrational contributions for disordered phases is computationally challenging, we relied on the observed trend that the relative free energies of the Li$_x$BC phases lie within $\sim$ 20 meV/atom to the LiBC$\leftrightarrow$BC tie-line.

\begin{figure}[t!]
\includegraphics[width=0.48\textwidth]{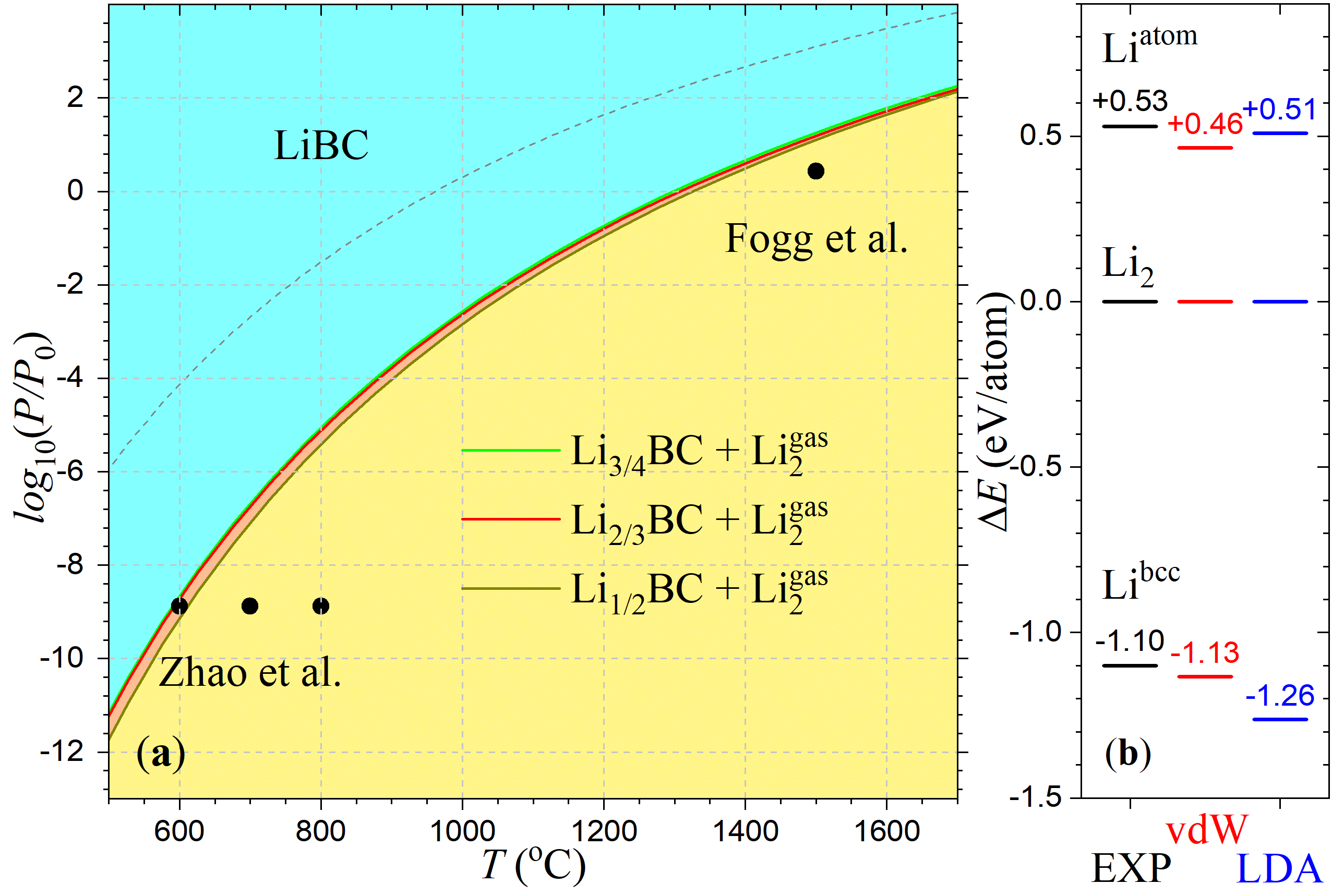}
\caption{(a) Calculated ($T$,$P$) phase boundary between ordered LiBC and Li$_{x}$BC+Li$_2^{\textrm{gas}}$ mixtures (solid lines) along with reported Li$_x$BC synthesis conditions (circles)~\cite{Zhao2003,Fogg2006}. The gray dashed line corresponds to a kinetically inaccessible phase transition between LiBC and C+B$_4$C+Li$_2^{\textrm{gas}}$. (b) Relative energy of Li referenced to the atomic energy of the Li$_2$ molecule, with experimental values taken from Refs. \cite{joseph2016} and \cite{hessel1979} (see Table S2 for more details).}
\label{fig:TP} \end{figure}

We estimated the location of the $x$-dependent phase boundaries using representative ordered oS44, hP8, and oI10 structures at $x$ = 3/4, 2/3, and 1/2, respectively. Once the vibrational entropy was included, we calculated the equilibrium $(T,P)$ values in the 800-2000 K range by matching the Gibbs free energies per Li atom for LiBC and the three mixtures of Li$_x$BC and Li$_2^{\textrm{gas}}$. The corresponding stability domains are plotted in Fig.~\ref{fig:TP}(a). We also considered a possible full decomposition of LiBC into Li$_2^{\textrm{gas}}$+C+B$_4$C (a slightly more stable combination than Li$_2^{\textrm{gas}}$+C+LiB$_6$C at elevated temperatures) shown with a gray dashed line. Thermodynamically, this transformation should happen first upon heating LiBC and/or reducing the Li$_2^{\textrm{gas}}$ vapor pressure and prevent the formation of any Li$_x$BC derivatives but, kinetically, it is evidently hindered by the high barriers associated with breaking the strong covalent B-C bonds. The presence of some metastable products makes the constructed phase diagram {\it transitional}\cite{dziubek2022}.

To gauge the sensitivity of the phase boundaries to systematic DFT errors, we compared the calculated and measured energies of the spin-polarized atomic, diatomic, and bcc Li (Fig.~\ref{fig:TP}(b)) \cite{boldyrev1993,joseph2016}. The slight overbinding of bcc-Li with respect to molecular Li$_2$ by 0.03 eV/Li in our default DFT approximation would result in a small shift in the boundary, $\Delta \log_{10}(P/P_0)\approx 0.03/0.07\log_{10}(e)\approx$ 0.19 at 800 K. Assessment of discrepancies between DFT and experiment for solid state phases is more challenging because measurements are usually conducted at elevated temperatures~\cite{ak47}. The typical differences of up to $\sim 20$ meV/atom between DFT flavors in the calculation of relative energies~\cite{ak28,ak47,ak48} and the neglected configurational entropy term of similar values (Fig.~\ref{fig:LixBC-dat}(b)) could be amplified to a sizable 0.2 eV/Li upon rescaling to be per Li atom. This could lead to up to an order of magnitude change in the estimate of the transition pressure.

We also performed QHA calculations to check how volume expansion affects the free energies of LiBC and Li$_{0.5}$BC at highest temperatures used in the delithiation experiments. As shown in Figs. S1  and S2, the minimum of $F(T,V)$ shifts by –21.53 meV/atom and –23.16 meV/atom at 1800 K, respectively, once the two phases are allowed to expand. This results in a –13.4 meV/Li change in the Gibbs free energy difference that defines the phase boundary between Li$_2$B$_2$C$_2$ and 1/2Li$_2$+LiB$_2$C$_2$. Other anharmonic terms for molecules and solids are much harder to evaluate from first principles~\cite{allen2020,monserrat2013,tadano2019}. Given the relatively small value of the QHA corrections compared to the uncertainties estimated above, the use of the harmonic approximation seems fitting in this case.

The reliability of the proposed phase diagram can be checked against explicitly specified synthesis conditions in two previous studies~\cite{Zhao2003,Fogg2006}. Zhao et al.~\cite{Zhao2003} carried out LiBC deintercalation at three relatively low temperatures maintaining a 10$^{-6}$ torr vacuum. The 12-hour experiments at 600, 700, and 800 $^\circ$C were estimated to reduce the Li content down to  $x=0.80$, 0.77, and 0.63, respectively. Fogg et al.~\cite{Fogg2006} performed deintercalation at higher $T=1500$~$^\circ$C and estimated the Li$_2^{\textrm{gas}}$ vapor partial pressure to be 2.7 atm using the Clausius-Clapeyron equation. Based on a series of experiments for durations between 1 and 24 hours which led to different $x$ values, the authors concluded that it is difficult to control the precise amount of Li in these dynamic non-equilibrium reactions.

\begin{figure*}[t!] 
\includegraphics[width=0.95\textwidth]{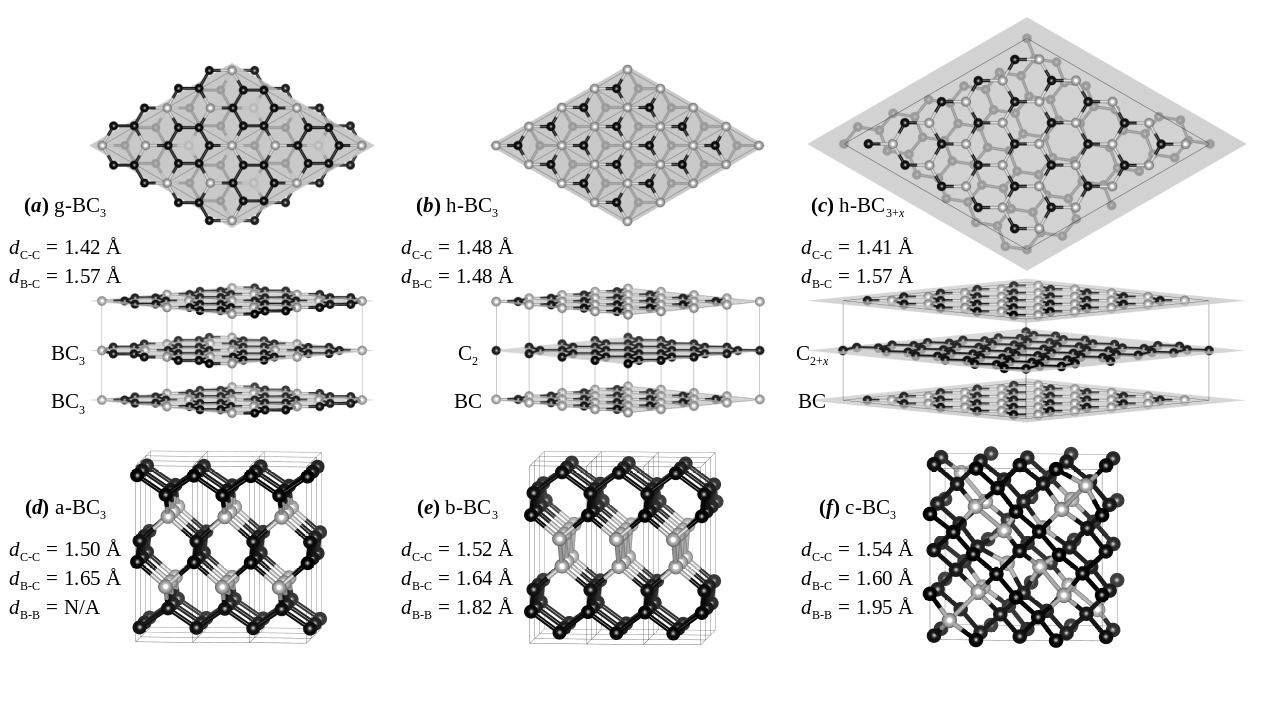} 
\caption{ Select structures with sp$^2$ (a-c) or sp$^3$ (d-f) bonding at or near the BC$_3$ composition. (a,b,f) Previously proposed models explaining experimentally observed phases. (c) A hybrid h-BC$_{3+x}$ unit cell with natural B-C and C-C bond lengths achieved by twisting BC and C$_2$ layers.}
\label{fig:BC3} \end{figure*}

The reported Li deintercalation $(T,P)$ conditions fall appropriately outside of our calculated LiBC stability domain (Fig.~\ref{fig:TP}(a)). The narrowness of the estimated stability regions for intermediate $x$ values suggests that the targeted $x$ values may indeed be easier to obtain by relying on the kinetics (e.g., tuning the duration and temperature profile of the deintercalation process) rather than thermodynamics (i.e., trying to pinpoint suitable $(T,P)$ conditions).

\subsection{BC$_3$}

Determination of the B-rich part of the B-C phase diagram has required several iterations~\cite{rogl2014,elliott1965,lowell1967,ploog1974,ploog1974-II, ploog1974crystal,aselage1991structural,aselage1992,kevill1986}. Originally identified as related but distinct hR45-B$_{13}$C$_2$~\cite{will1979} and hR45-B$_4$C~\cite{clark1943} compounds with the $R \overline{3}m$ symmetry, they have been ultimately found to be a single B$_{4+\delta}$C intermediate phase with a large homogeneous region of ($0.088 < x \leq 0.2$) in B$_{1-x}$C$_{x}$. The composition variability arises from the thermodynamic favorability of connecting the B$_{12}$ icosahedra with different blocks \cite{morosin1986, lazzari1999, mauri2001, calandra2004}. In our study, we use B$_4$C \cite{bylander1990,lazzari1999,mauri2001}, B$_{6.67}$C \cite{jay2019}, and B$_{10.5}$C \cite{jay2019} models that fall into the stability range at low temperatures. The first one was constructed as CBC-B$_{11}$C$^{\textrm{p}}$ with a C atom populating a polar site in B$_{12}$ and a B atom populating the mid-point site in three-atom chains shown in Fig.~\ref{fig:all}. The last two B-rich representative structures were obtained by expanding the hR45-B$_{13}$C$_2$ primitive unit cell and distributing B$\langle_{\textrm{B}}^{\textrm{B}}\rangle$B (OPO1) or CBC...B (OPO2) blocks in the $3\times 3\times 3$ supercell as described in Ref.~\cite{jay2019} (see Supporting Information for more details). 
We note that the tie-lines defined by the B$_{4+\delta}$C phase have little bearing on the stability of the BC$_3$ polymorphs known to have large positive formation energies \cite{liu2011,jay2019}.

The feasibility of synthesizing a borocarbide at the 1:3 composition was demonstrated by Kouvetakis et al. in 1986~\cite{kouvetakis1986}. A metastable graphite-like g-BC$_3$ phase was obtained via a chemical vapor deposition reaction of 2BCl$_3$ with C$_6$H$_6$ at 800 $^\circ$C yielding 2BC$_3$ and 6HCl. Further experimental work has refined the compound’s structure with electron energy-loss spectrum analysis~\cite{krishnan1991} and demonstrated a way of producing the material in large quantities via thermolysis of the aromatic boron compound 1,3-bis(dibromoboryl)benzene~\cite{king2015}. The results indicate that g-BC$_3$ features a uniform ordering of B atoms within turbostratically disordered layers~\cite{king2015}. 

Sun et al.~\cite{sun2004} found g-BC$_3$ (Fig.~\ref{fig:BC3}(a)) to be dynamically unstable in LDA calculations and constructed several more stable, up to 22 meV/atom, derivatives. We also observed a full phonon branch along $\Gamma$-A to have imaginary frequencies and examined the stability and symmetry of structures derived from g-BC$_3$ (see Fig. S3). An interlayer shift leading to an orthorhombic e-BC$_3$ polymorph with an oS32 (Fmmm) unit cell lowered the energy by 15 meV/atom and eliminated the dynamical instability at $\Gamma$ but left imaginary frequency modes at the Y point in our optB86b-vdW simulations (Fig. S3). By randomly distorting g-BC$_3$ and performing full unit cell optimizations, we obtained a low-symmetry aP16 (P$\overline{1}$) variant, r-BC$_3$, 3 meV/atom below e-BC$_3$ but still dynamically unstable. It is evident that the energy landscape has numerous nearly degenerate minima corresponding to different layered sequences with large unit cells, which is consistent with the experimentally observed stacking disorder.

The discovery of the layered g-BC$_3$ has offered possibilities of creating a MgB$_2$-type hole-doped superconductor~\cite{ribeiro2004} or a battery anode for high-capacity Li-ion storage~\cite{kuzubov2012,liu2013,king2015,joshi2015}. Given a well-known pressure-induced sp$^2$ to sp$^3$ transformation pathway in pure carbon, g-BC$_3$ was also proposed to serve as a precursor for making a superhard t-BC$_3$ diamond analog, shown to be enthalpically favored above 4 GPa~\cite{Liu2006}. Zinin et al. did obtain a dense BC$_3$ polymorph by heating the starting layered material to 2033~K at 50~GPa~\cite{zinin2007} but its precise structure proved difficult to determine. Several low-enthalpy decorations of the diamond structure were identified with particle swarm optimizations in follow-up studies~\cite{liu2011,zhang2015}, such as (a-c)-BC$_3$ shown in Fig.~\ref{fig:BC3}(d-f). Simulated x-ray diffraction (XRD) and Raman peaks for the lowest-enthalpy cubic c-BC$_3$ phase at $\sim 40$~GPa were shown to be in excellent agreement with the experimental data~\cite{zhang2015}. It is worth noting that none of the polymorphs become truly thermodynamically stable under compression, e.g., their formation enthalpies with respect to $\gamma$-B and diamond-C are over +160 meV/atom at 40 GPa. At ambient pressure, a-BC$_{3}$ has the lowest energy among all configurations considered in previous and present studies (see Table S1).

In 2018, Milashius et al. reported the synthesis of a novel hybrid layered BC$_3$ phase (h-BC$_3$) comprised of alternating carbon and borocarbide honeycomb layers~\cite{milashius2018}. The samples were prepared by heating pure elements to 1200 $^\circ$C followed by rapid cooling and annealing at 400 $^\circ$C. Using powder XRD and the Rietveld method, the authors described the material with an ordered $P\overline{6}$m2 hP4 structure shown in Fig.~\ref{fig:BC3}(b). The following examination highlights a few inconsistencies of the proposed model. 

\begin{figure}[t!]
\includegraphics[width=0.45\textwidth]{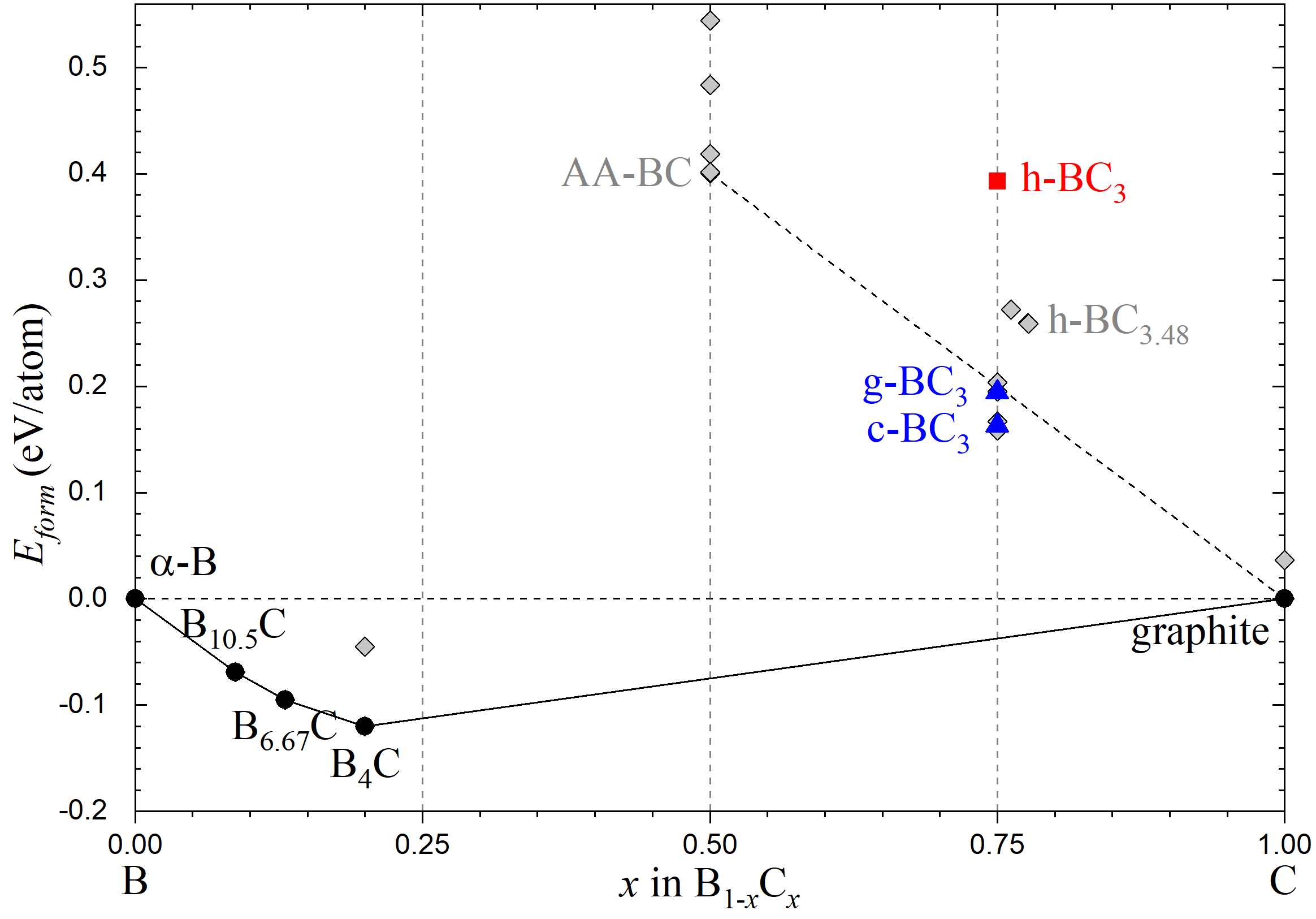}
\caption{Calculated formation energies for considered B-C phases. The known B$_{4+\delta}$C is represented with three related B-rich stable structures (black circles). The reported BC$_3$ polymorphs are divided into metastable (blue triangles) and unstable (red square). Hypothetical unstable phases, e.g., the layered AA-BC and h-BC$_{3.48}$, are included for reference (gray diamonds).}
\label{fig:BC} \end{figure}

First, h-BC$_3$ has a large positive formation energy of over 400 meV/atom. To put it into context, we simulated different stackings of ordered 1:1 borocarbide layers and plotted a tie-line connecting the most favorable, albeit dynamically unstable, AA configuration and graphite in Fig.~\ref{fig:BC}. The closeness of the metastable g-BC$_3$ formation energy to the interpolated value reflects the structure’s capacity to accommodate the B-C and C-C bonds with near-optimal lengths, 1.57 \AA\ and 1.42 \AA\ obtained for BC and graphite, respectively (Fig.~\ref{fig:BC3}(b)). In contrast, h-BC$_3$ constraints these covalent bonds to a common intermediate value of 1.48 \AA. Our results in Fig. S4 illustrate that a substantial portion of the excess 200 meV/atom corresponds to the elastic energy stored in the compressed B-C and stretched C-C bonds. To illustrate that most of the energy could be released without breaking the covalent bonds within the bulk of h-BC$_3$, we constructed related hybrid structures with natural bond lengths by combining rotated graphite and heterographite supercells. The hP112 structure shown in Fig.~\ref{fig:BC3}(c) with $(6\vec{a}+\vec{b};-\vec{a}+5\vec{b})$ and $(5\vec{a};5\vec{b})$ lateral expansions of the C and BC hexagonal two-atom unit cells ensures a match within 0.15\% between the two unrelaxed sublattices, which makes the resulting off-stoichiometric h-BC$_{3+0.48}$ phase significantly less unstable (Fig.~\ref{fig:BC}). In addition, hBC$_3$ was found to be dynamically unstable and feature multiple modes with imaginary frequencies, e.g,. $34i$ cm$^{-1}$ at $\Gamma$ and $396i$ cm$^{-1}$ at M points. These findings indicate that the h-BC$_3$ formation from pure B and C would not be favored either thermodynamically or kinetically.

\begin{figure}[t!] 
\includegraphics[width=0.48\textwidth]{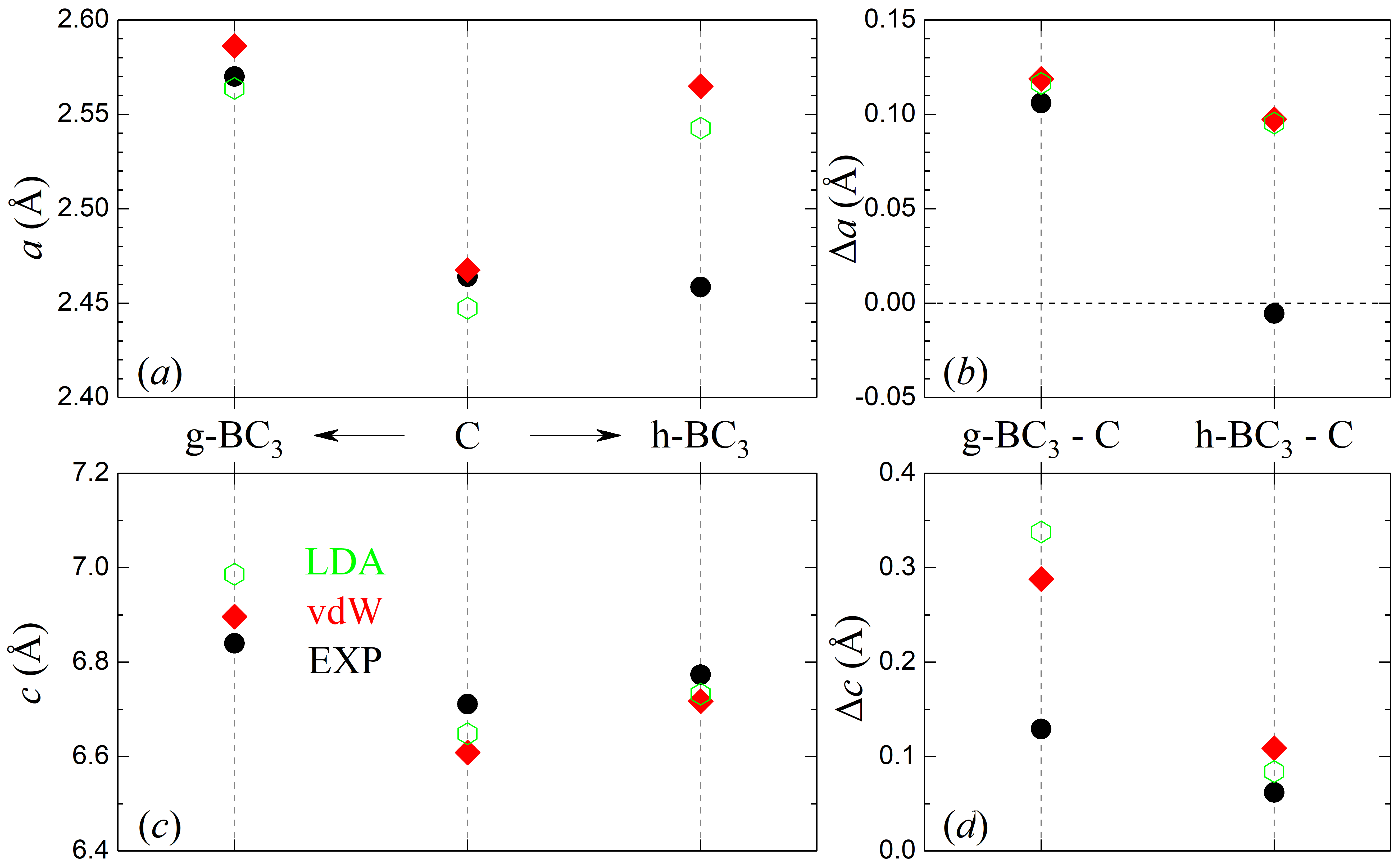}
\caption{Comparison of lattice constants in graphite and BC$_3$ polymorphs extracted from XRD data and evaluated with different DFT functionals in this work. The experimental values are taken from Refs.~\cite{king2015, ueno2006} for g-BC$_3$ and from Ref.~\cite{milashius2018} for h-BC$_3$ and graphite. Panels (b,d) show change in the lattice constants in g-BC$_3$ and h-BC$_3$ with respect to those in graphite.} \label{fig:BOND} \end{figure}

Second, the 2.4586 \AA\ in-plane lattice constant extracted from powder XRD data for h-BC$_3$ appears to be unphysically small, given that the corresponding value for graphite is 2.466 \AA~\cite{milashius2018}. Substitution of a quarter of C atoms for larger B atoms should lead to a noticeable expansion of the lattice, regardless of whether the minority species is distributed in every (g-BC$_3$) or every other (h-BC$_3$) layer. We compare the lattice constant change relative to pure graphite in Fig.~\ref{fig:BOND} to reduce the value of known 1-2\% systematic DFT errors in the evaluation of bond lengths. According to our optB86b-vdW and LDA results, the in-plane dimensions in both borocarbide structures expand by about 4\% reaching 2.586 \AA\ in g-BC$_3$ with the former functional. Unfortunately, no $(hk0)$ peaks have been observed in g-BC$_3$ bulk samples due to stacking disorder~\cite{king2015} and the only information about the in-plane dimensions was obtained with scanning tunneling microscopy for BC$_3$ monolayers on NbB$_3$(0001)~\cite{ueno2006,yanagisawa2006}. The non-strained 2D monolayer (denoted as s-BC$_3$ in Ref.~\cite{yanagisawa2006}) incommensurate with the underlying substrate was found to have $a_{\textrm{C-C}}=1.42$ \AA, $a_{\textrm{B-C}}=1.55$ \AA, and $a=2.57$ \AA\ consistent with our optimized values. The interlayer distances defined by the weak dispersive interactions are harder to reproduce with (semi)local DFT approximations~\cite{ak06,graziano2012,lenchuk2019} and match experimental values within 1-2\% with the employment of vdW functionals. The elongation of the $c$-axis to 6.77 \AA\ in h-BC$_3$ noted in Ref.~\cite{milashius2018} is actually less pronounced than the measured and calculated values in g-BC$_3$ (Fig.~\ref{fig:BOND}(d)). In fact, turbostratic forms of graphite obtained via heat treatment or ball milling have been reported to have larger $c$-axis values reaching 6.88 \AA \cite{bayot1990,li2007}. Fig. S5 compares the positions and shapes of the key powder XRD peaks observed for the layered carbon and borocarbide materials. The h-BC$_3$ pattern~\cite{milashius2018} stands out in that it has relatively sharp peaks, includes reflections dependent on the in-plane lattice constant, and contains no significant signal from ordered graphite.

We conclude that while the layered h-BC$_3$ model proposed by Milashius et al. provides a reasonable fit to their powder XRD data and is consistent with the 1:3 composition established with a microprobe analysis~\cite{milashius2018}, it is deficient from the thermodynamic or chemical bonding points of view. More experimental information is needed to solve the intriguing new material prepared at this composition.

\subsection{LiBC$_3$}

Borocarbides with lighter, larger, electron-deficient networks have been considered as promising alternatives to graphite for Li-ion battery applications. Intercalation of boron-substituted B$_{z}$C$_{1-z}$ graphite ($z\approx 0.16$) and deintercalation of the stoichiometric LiBC have indeed offered an increased reversible specific capacity of 437 mAh/g and 450 mAh/g, respectively~\cite{way1994,li2018}. First-principles calculations showed that g-BC$_3$ could accommodate up to Li$_{1.5}$ per formula unit, which corresponds to a theoretical storage capacity of 857 mAh/g~\cite{liu2013,joshi2015}. King et al.’s experimental study on the g-BC$_3$ intercalation observed a 700 mAh/g irreversible Li uptake upon the first charge and a 374 mAh/g cycling reversibility, presumably between Li$_{0.65}$BC$_3$ and Li$_{1.5}$BC$_3$ compositions~\cite{king2015}.

\begin{figure}[t!]
\includegraphics[width=0.48\textwidth]{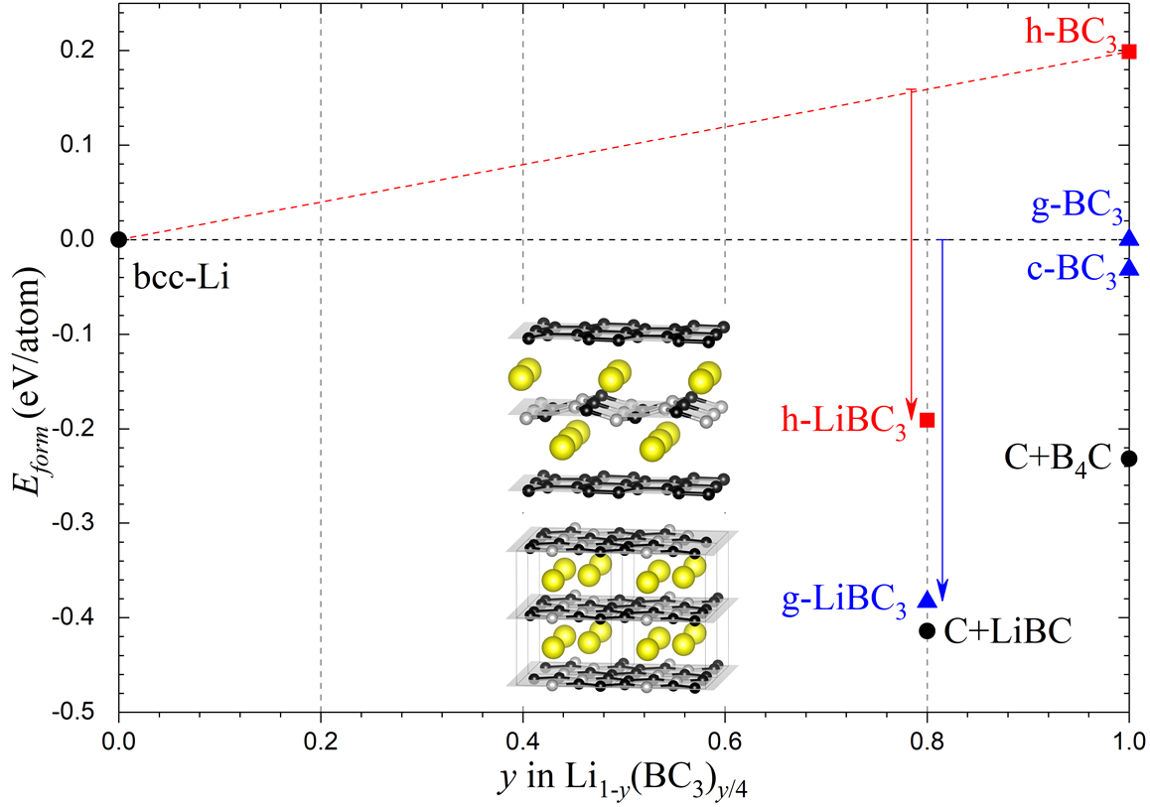}
\caption{Relative energies of intercalated LiBC$_3$ phases with respect to bcc-Li and g-BC$_3$. The solid points at $y=0.8$ and $y=1.0$ represent the convex hull energies for 2/5C + 3/5LiBC and 11/16C+5/16B$_4$C mixtures, respectively.}
\label{fig:LixBC3} \end{figure}

The LiBC$_3$ compound synthesized by Milashius et al. directly from the elements was determined with powder and single-crystal XRD measurements to have a hexagonal unit cell with the $P\overline{6}m$2 symmetry and an ordered distribution of atoms in alternating C-C and B-C layers~\cite{milashius2017}. We approximated the reported lithium borocarbides featuring partial occupation of Li sites~\cite{king2015,milashius2017} with small ordered g-LiBC$_3$ (oP10) and h-LiBC$_3$ (oP10) structures shown in Fig.~\ref{fig:LixBC3} (for comparison, a hexagonal hP15 polymorph was found to be 24 meV/atom less stable than the orthorhombic one for h-LiBC$_3$).

The relative stability results shown in Fig.~\ref{fig:LixBC3} reveal that the g and h forms of LiBC$_3$ are below the corresponding Li$\leftrightarrow$g-BC$_3$ and Li$\leftrightarrow$g-BC$_3$ tie lines by similar 0.38 eV/atom and 0.35 eV/atom margins, respectively. While the Li intercalation brings g-LiBC$_3$ close to true thermodynamic stability (within 31 meV/atom), it does not help h-LiBC$_3$ overcome the energy penalty inherited from the h-BC$_3$ parent structure. The oP10 model of h-LiBC$_3$ is also dynamically unstable. Therefore, our conclusions on the thermodynamic and kinetic feasibility of the hybrid morphology for BC$_3$ apply for the lithiated derivatives as well. The disagreement between experimental (2.5408 \AA) and DFT (2.5981 \AA) in-plane lattice constants in h-LiBC$_3$ is less pronounced (2.2\%) than in h-BC$_3$ (4.1\%, see Fig.~\ref{fig:BOND}), but still at the higher end of typical systemic DFT errors.

\subsection{Li$_2$B$_2$C}
\begin{figure}[t!]   
  \includegraphics[width=0.48\textwidth]{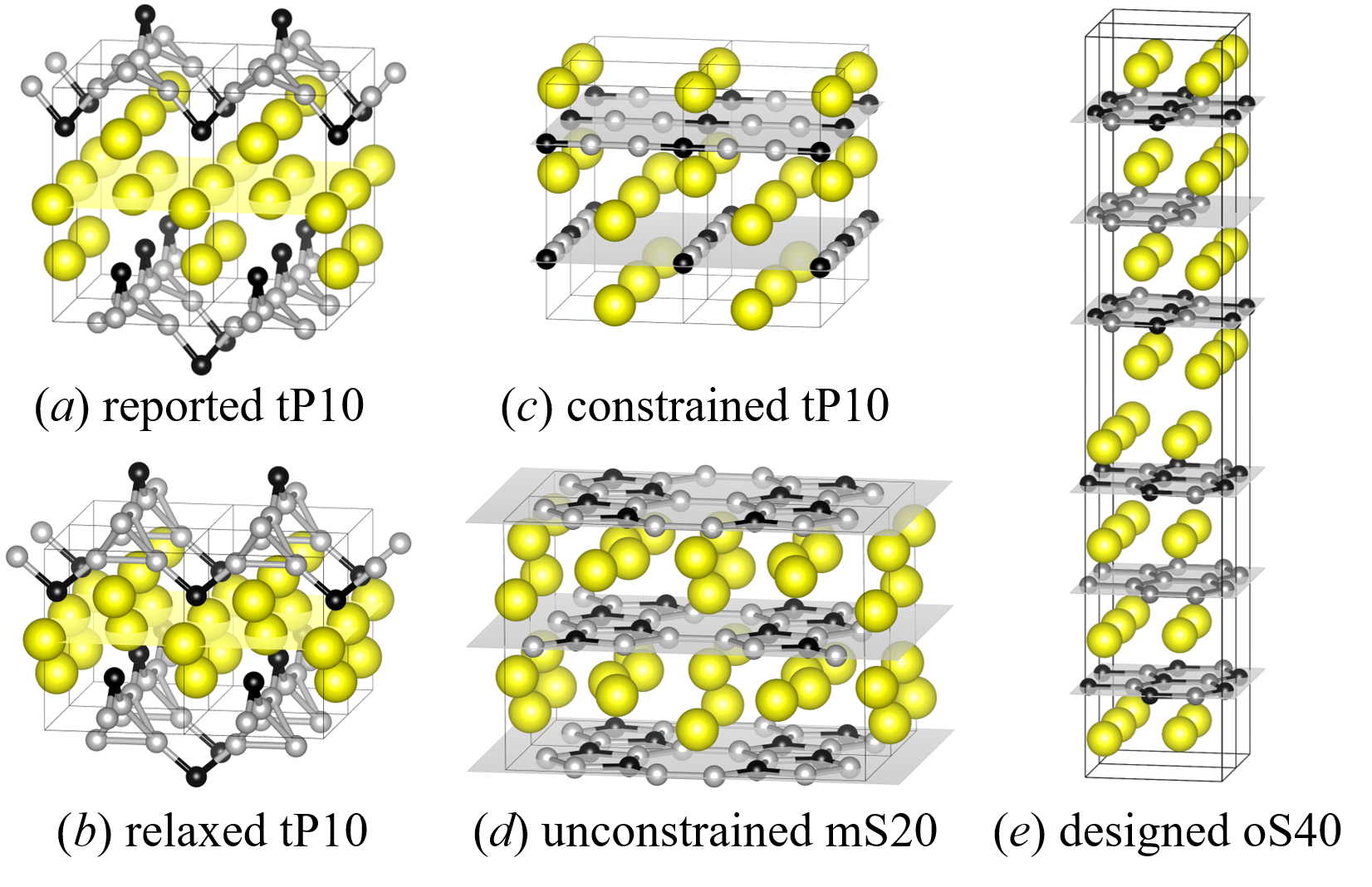} \caption{ 
   Crystal structures of select Li$_2$B$_2$C phases labeled with Pearson symbols. The Li, B, and C atoms are shown with yellow, gray, and black spheres, respectively. The methods used to construct the polymorphs are detailed in the main text and Fig.~\ref{fig:Li2B2C}.}
\label{fig:Li2B2C-pic} 
\end{figure}

Pavlyuk et al. observed the formation of a new Li$_2$B$_2$C compound after a mixture of the elements at the nominal composition was heated at 1473 K and rapidly cooled to room temperature~\cite{pavlyuk2015}. Collected powder XRD results did not match patterns of any known Li-B-C phases. Based on single-crystal data, the authors proposed a tetragonal structure with space group $P\overline{4}m$2 shown in Fig.~\ref{fig:Li2B2C-pic}(a). However, our analysis of the thermodynamic stability and chemical bonding of tP10-Li$_2$B$_2$C reveals several issues with the reported solution.

First, the local relaxation of the original structure changed the base lattice constants by less than 0.4\% but resulted in a dramatic 25\% collapse of the $c$-axis, from 7.1055 \AA\ to 5.342 \AA. While large interlayer spacing mismatches between experiment and (semi)local DFT approximations have been observed for van der Waals solids~\cite{ak06,graziano2012,lenchuk2019}, the comparable results obtained in the LDA (22\%) and PBE (25\%) treatments make the systematic DFT errors an unlikely source for the discrepancy. Moreover, the 2.6 eV/Å starting atomic forces and the ensuing 0.33 eV/atom stabilization are too large to be attributed to the typical 1-2\% DFT systematic errors for covalent or metallic bond lengths.

Pavlyuk et al. made an interesting comparison between the connectivity of building blocks in the proposed tP10-Li$_2$B$_2$C and in several known compounds~\cite{pavlyuk2015}. However, the 2D display of the atomic arrangements and the reference to the units as B$_4$ and B$_2$C$_2$ squares do not reflect the true 3D connectivity of the covalent frameworks in the YB$_2$C$_2$, CrB$_4$~\footnote{Ref.~\cite{pavlyuk2015} displayed a previously misidentified oI10 crystal structure for CrB$_4$. The compound was predicted~\cite{ak17} and confirmed~\cite{ak22} to have a lower-symmetry oP10 ground state structure with a significantly distorted 3D boron framework.}, CeB$_2$C$_2$, LiB$_3$, and ThB$_4$ compounds featuring octahedra and diamond-like local environments. In fact, a visual examination of the original tP10 structure in  Fig.~\ref{fig:Li2B2C-pic}(a) reveals that the C and B atoms in the corrugated networks are actually undercoordinated, having only two and three neighbors within 1.9 \AA, respectively. The local optimization brings an additional neighbor within the effective interaction range for B but cannot improve the atypical local environment with a dangling bond for C (Fig.~\ref{fig:Li2B2C-pic}(b)). It leaves the structure 0.437 eV/atom above the convex hull (Fig.~\ref{fig:LixBC-dat}) and dynamically unstable, with imaginary frequencies as high as $198i$ cm$^{-1}$ across the Brillouin zone.

Next, we attempted to find more stable Li$_2$B$_2$C polymorphs using our evolutionary algorithm. We started by constraining our searches to the reported unit cell dimensions and the $P\overline{4}m$2 space group. The inclusion of prior information extracted from experiment is known to accelerate the identification of ground states by 2-3 orders of magnitude~\cite{ak23}. These constrained evolutionary runs did produce an alternative tP10 structure with lower energy (by 66 meV/atom after full unit cell relaxation), higher symmetry ($P4_2/mmc$), and conspicuously different morphology with crisscrossing B$_2$C linear chains (Fig.~\ref{fig:Li2B2C-pic}(c)). Given the difficulty of determining the exact atomic positions of light elements in XRD measurements and the significant reduction of volume upon relaxation of the original tP10 structure, we also tried to fit three Li$_2$B$_2$C formula units (15 atoms) into the reported unit cell but the best candidate turned out to be suboptimal. Additional runs with 10-15 atoms per unit cell and compositions near 2:2:1 did not produce viable Li-B-C phases with the $P\overline{4}m$2 symmetry.

\begin{figure}[t!]   
  \includegraphics[width=0.48\textwidth]{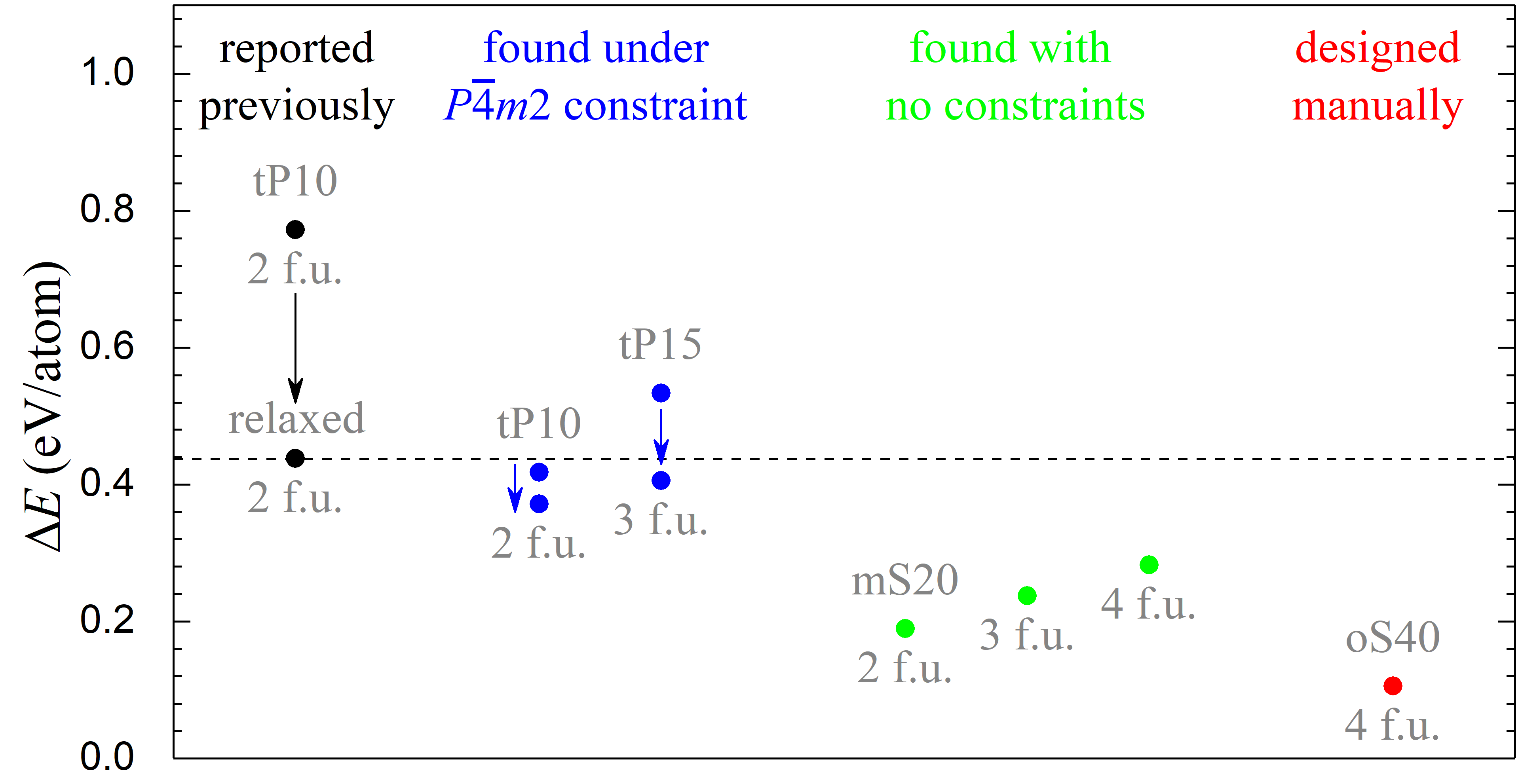} \caption{ 
   Distance to the convex hull for Li$_2$B$_2$C competing structures reported previously~\cite{pavlyuk2015} and identified in this study. The size of the unit cell is specified with the number of formula units (f.u.). The arrows indicate the energies before and after the full relaxation of the lattice parameters.}   
\label{fig:Li2B2C} 
\end{figure}

We proceeded with unconstrained evolutionary searches initialized with random configurations. The global optimization uncovered substantially more stable polymorphs. Runs with two formula units yielded the best mS20 structure with an interesting layered morphology. As can be seen in Fig.~\ref{fig:Li2B2C-pic}(d), the three-fold connectivity satisfying the Euler’s rule~\cite{ak12} is maintained via the combination two pentagons and one octagon rather than three hexagons. The B$_2$C framework enabling sp$^2$ bonding makes mS20 more stable by over 200 meV/atom compared to the other candidates (Fig.~\ref{fig:Li2B2C}). Searches with larger system sizes did not produce better configurations; the fact that our extensive runs with four formula units did not reproduce the energy of the best candidates with two formula units highlights the difficulty of finding complex stable configurations with 20 atoms.

Finally, we relied on general knowledge of favorable motifs in M-B and M-B-C materials to manually construct better polymorphs. The title compound can be thought of a combination of (LiB)$_2$ and C materials known to be stable as stackings of honeycomb covalent networks~\cite{ak08}. As discussed in the case of BC$_3$, the B- and C-based layers are too different in size to be combined as separate units. Mixing B and C within the LiB structure would not maintain the desired composition. Therefore, we examined a hybrid LiB-MgB$_2$ structure introduced in our previous study~\cite{ak48} and sampled different decorations of the B/C sites. The resulting oS40 structure shown in Fig.~\ref{fig:Li2B2C-pic}(e) proved to be a significantly more stable configuration. Because of the large $c/a$ ratio, standard evolutionary operations could not create it. Despite being our best guess at this composition, oS40 remains 105 meV/atom above the convex hull (Fig.~\ref{fig:Li2B2C}) and is not expected to form.

\section{Conclusions}

The presented ab initio analysis further elucidates the interplay between thermodynamic and kinetic factors governing the formation of Li-B-C compounds. Whereas the observed delithiated Li$_x$BC ($x>0.38$) phases and BC$_3$ (g and c) polymorphs are known to be only metastable under ambient conditions, our results indicate that they are metastable under synthesis conditions as well. In case of Li$_x$BC, the delithiation process is driven by the high entropy of Li$_2^{\textrm{gas}}$ at elevated temperatures while the strongly bonded BC layers prevent the material’s decomposition into other products, e.g., Li$_2^{\textrm{gas}}$+C+B$_4$C. Our constructed ab initio $(T,P)$ phase diagram is in good agreement with previous observations~\cite{Zhao2003,Fogg2006}. For BC$_3$, the synthesis of sp$^2$ and sp$^3$ compounds relies on natural rebonding pathways between precursor and targeted materials. On the other hand, reported h-BC$_3$, h-LiBC$_3$, and Li$_2$B$_2$C phases obtained from the elements ~\cite{pavlyuk2015,milashius2017,milashius2018} have been shown to have unnatural structural features and high positive formation energies. We hope that our findings will stimulate further synthesis and characterization work on this interesting materials class.

\section*{Conflicts of interest}
There are no conflicts to declare.

\section*{Acknowledgements} 
The authors acknowledge support from the National Science Foundation (NSF) (Awards No. DMR-2132586 and No. DMR-2132589). This work used the Expanse system at the San Diego Supercomputer Center via allocation TG-DMR180071 and the Frontera supercomputer at the Texas Advanced Computing Center via the Leadership Resource Allocation (LRAC) grant No. 2103991 (allocation DMR22004). Expanse is supported by the Extreme Science and Engineering Discovery Environment (XSEDE) program~\cite{xsede} through NSF Award No. ACI-1548562, and Frontera is supported by NSF Award No. OAC-1818253~\cite{Frontera}.





\bibliography{lit} 
\bibliographystyle{apsrev4-2} 

\end{document}